\documentstyle[amsmath,aps,prc,epsf,floats,twocolumn]{revtex}

\begin{document}
\draft
\twocolumn[\hsize\textwidth\columnwidth\hsize\csname @twocolumnfalse\endcsname

\title{J/$\psi$ suppression in heavy ion collisions
by quark momentum diffusion}
\author{H.\ Fujii}
\address{Institute of Physics, University of Tokyo\\
3-8-1 Komaba, Meguro, Tokyo 153-8902}
\date{\today}
\maketitle

\begin{abstract}
The momentum diffusion effect of the quark pair
due to the multiple scattering in a nuclear medium
is studied to explain 
the observed J/$\psi$ yields in SPS experiments.
The resulting suppression is found to be insufficient to reproduce
the J/$\psi$ yield in Pb-Pb collisions at SPS energy.
\end{abstract}
]

The suppression of the J/$\psi$ cross section
in ultra-relativistic nucleus-nucleus
collisions has drawn much attention since its proposal as
a possible signal of the quark-gluon plasma (QGP)\cite{MS86},
and the interest is even more increased
after the anomalous suppression observed 
in Pb-Pb interactions at SPS energy~\cite{NA50}. 
There have been so many works done for explaining
this anomalous suppression
in the conventional manner or in the QGP-presuming
 way~\cite{RV99,HS00}.

Qiu, Vary and Zhang (QVZ) recently proposed a model~\cite{QVZ02}
 for calculating
the J/$\psi$ suppression caused by the nuclear effect
in high-energy proton-nucleus (p$A$) and nucleus-nucleus ($AB$)
collisions.
Their model is based on the QCD factorization of 
the inclusive processes. The nuclear medium effect
is taken into account as a modification of the transition probability 
of the $c\bar c$ pair due to the multiple scattering in the 
nuclear environment.
The multiple scattering effects have been observed as the momentum
imbalance of two jets in hadron-nucleus 
collisions~\cite{LQG94,XG98}.
For the J/$\psi$ production in heavy ion collisions,
the multiple scattering will increase the relative momentum of the pair
and accordingly push the invariant mass toward
the open charm threshold
as it traverses the nuclear medium.
QVZ model successfully explains
the observed suppression from p$A$ to $AB$
collisions including the Pb-Pb data
as nuclear effect~\cite{QVZ02,AKC02}.

In QVZ model, the net effect of the multiple scattering 
is expressed as a shift of the relative momentum of the quark 
pair in the transition probability function(see Eq.\ (\ref{eq:shift})).
This shift operator may be obtained by summing up all twist contributions
at the leading order in the strong coupling constant $\alpha_s$
and the nuclear size, in the perturbative QCD calculation\cite{F02,BQV94}.
This is quite different from the result of the
classical multiple scattering process;
after independent multiple collisions with random momentum transfers,
the distribution of the relative momentum $q$ of the pair diffuses while the
mean $q^2$ increases.
The only aim of this Note is to examine to what extent
the J/$\psi$ yield in heavy ion collisions is explained
as the influence of the momentum diffusion,
treating the multiple scattering of the quark pair as random walk.
We show that the resulting suppression is insufficient to explain
the Pb-Pb data and looks rather similar to the Glauber model result.

Let us first review QVZ model\cite{QVZ02}.
The inclusive cross section of the J/$\psi$ production in the
collision of the hadrons $A$ and $B$ is written in 
a factorized form:
\begin{eqnarray}
&&\sigma_{AB\to {\rm J}/\psi X}=
K_{{\rm J}/\psi}\sum_{a,b}
\int d q^2 \left ( \frac{
\hat \sigma_{ab\to c \bar c}(Q^2)}{Q^2}\right )
\nonumber\\
&& \times
\int dx_F \phi_{a/A}(x_a)\phi_{b/B}(x_b)
\frac{x_a x_b}{x_a+x_b}F_{c\bar c\to {\rm J}/\psi}(q^2),
\label{eq:xsec}
\end{eqnarray}
where $\sum_{a,b}$ runs over all parton flavors,
 $Q^2=q^2+4 m_c^2$, $\phi_{a/A}(x_a)$ is the distribution function
of parton $a$ in hadron $A$, and
$x_F=x_a-x_b$ and $x_a x_b=Q^2/s$.
The parton cross section $\hat \sigma$ is given 
in \cite{BQV94}.
This is the leading order formula in 
$\alpha_s$ and the phenomenological constant $K_{{\rm J}/\psi}$
corrects the higher order effects.
$F_{c\bar c\to {\rm J}/\psi}(q^2)$ describes the transition probability for
the $c\bar c$ state of the relative momentum $q^2$ to evolve into
a physical J/$\psi$ meson.
For this transition probability they propose a parametrization of 
\begin{eqnarray}
F^{(P)}_{c\bar c\to {\rm J}/\psi}(q^2)
&=&N_{{\rm J}/\psi}\theta(q^2)\theta(4{m'}^2-4m_c^2-q^2)
\nonumber \\
&\times&
\left (1-\frac{q^2}{4{m'}^2-4m_c^2}
\right )^{\alpha_F}  ,
\label{eq:powerfrag}
\end{eqnarray}
which includes the effect of the open charm threshold
at $4{m'}^2$ and simulates the gluon radiation effect with the parameter 
$\alpha_F>0$ with putting the larger weight to the smaller $q^2$.
The constant probability ($\alpha_F=0$)
corresponds to the color-evaporation (CE) model.

For the J/$\psi$ production in the p$A$ and $AB$ collisions 
QVZ model additionally assumes the separation of 
the multiple scattering which affects the pair state
and the formation of the J/$\psi$ resonance in the high-energy
interactions.
The multiple scattering of the pair in the nuclear medium
would increase the relative momentum $q^2$
of the pair.
The effect of coherent multiple scattering 
in the perturbative QCD calculation
may be represented by shifting of the relative momentum 
in the transition probability \cite{QVZ02} as 
\begin{equation}
F_{c\bar c\to {\rm J}/\psi}(\bar q^2)
=F_{c\bar c\to {\rm J}/\psi}(q^2+\varepsilon^2 L),
\label{eq:shift}
\end{equation}
where $L$ is the effective length of the nuclear medium
in the $AB$ collisions.
We note here that for a large enough $L$ such that
$\bar q^2>4{m'}^2-4m_c^2$
the transition probability
essentially vanishes due to the existence of the open charm
threshold~(\ref{eq:powerfrag}).
This apparently gives rise to a 
much stronger suppression than the exponential one
following from the Glauber model.

We would like to describe now 
the momentum diffusion effect by modifying the transition probability
$F_{c\bar c\to {\rm J}/\psi}$ of QVZ model.
The $c\bar c$ pair
with relative momentum, $q$,
produced in a hard parton collision, will change its momentum
to $q'$ after the random multiple scattering, and then transforms into
the J/$\psi$ with the probability $F_{c\bar c\to {\rm J}/\psi}({q'}^2)$.
Here we treat the exchanged soft momenta as independent and random
in the multiple scatterings.
After many scatterings, this classical, elementary diffusion process 
of the momentum results 
in  the Gaussian distribution around the initial value $q$ 
with the variance $\varepsilon ^2 L$.
The three-dimensional random walk is assumed here only for simplicity.
The transition probability should be in effect
replaced by the one smeared with this Gaussian weight as
\begin{equation}
\bar F_{c\bar c\to {\rm J}/\psi}(q^2)
 \equiv \frac{1}{(2\pi \varepsilon^2 L)^{3/2}}
\int d^3 q' e^{- \frac{(q'-q)^2}{2\varepsilon^2 L}}
F_{c\bar c\to {\rm J}/\psi}({q'}^2).
\label{eq:smear}
\end{equation}
Note that
the transition probability $\bar F_{c\bar c\to {\rm J}/\psi}(q^2)$
never vanishes for any $q$ although
the average momentum of the pair increases
as $\langle {q'}^2\rangle=q^2+3 \varepsilon^2 L$
 after the multiple scattering.
We immediately find that 
the transition probability behaves for the asymptotically large $L$ 
as
\begin{equation}
\bar F_{c\bar c\to {\rm J}/\psi}(q^2) \sim
 \frac{1}{(2\pi \varepsilon^2 L)^{3/2}}
\int d^3 q'F_{c\bar c\to {\rm J}/\psi}({q'}^2).
\label{eq:asympt}
\end{equation}
This power suppression in $L$ which  stems from the
depletion of the normalization factor
is more moderate than the exponential one.

\begin{figure}
\centerline{\epsfxsize=0.38\textwidth \epsffile{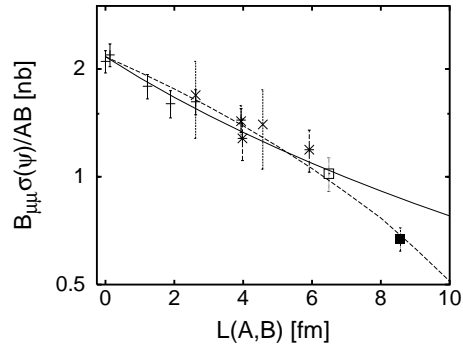}}
\caption{\label{fig1} J/$\psi$ cross section to
$\mu^+ \mu^-$ pair\protect\cite{NA50}
compared with the models using the smeared
transition probability Eq.~(\protect\ref{eq:smear}) (solid line) 
and the transition probability with the shifted momentum
Eq.~(\protect\ref{eq:shift})
(dashed line). }
\end{figure}

In Fig.~\ref{fig1}
we show our result on the J/$\psi$ suppression 
calculated using the formula (\ref{eq:xsec})
with the smeared probability (\ref{eq:smear}).
The parton distribution function of CTEQ5L\cite{CTEQ5}
is used without any nuclear
modification, and the parameters are fixed to the same as \cite{QVZ02}:
$\alpha_F=1$ and $f_{{\rm J}/\psi}\equiv 
K_{{\rm J}/\psi} N_{{\rm J}/\psi}=0.485$ for the 
transition probability (\ref{eq:powerfrag}). 
The multiple scattering is included as smearing with
$\varepsilon^2=0.185$ GeV$^2$/fm.
Our model reasonably fits the data in the p$A$ and $AB$
collisions taken from \cite{NA50}
except the Pb-Pb point. The curve bends upward in the semi-log plot
as is expected from Eq.~(\ref{eq:asympt}), but is almost consistent
with the straight line within this interval of $L$.
The original QVZ model (\ref{eq:shift})
with $\varepsilon^2=0.25$ GeV$^2$/fm 
(dashed line) can explain  all the data points in Fig.~\ref{fig1}.
The downward bending of QVZ model 
is the result of the existence of the open charm threshold in the
transition probability~(\ref{eq:powerfrag}) and the uniform shift
of the momentum~(\ref{eq:shift}).

We performed the calculations using other forms
for the transition probability,
the Gaussian form and the CE form,
besides Eq.~(\ref{eq:powerfrag})\cite{QVZ02},
and confirmed that the qualitative behavior of the suppression is unaltered;
the Pb-Pb data alone lies far below the curve calculated by our model.
The concavity of the suppression curve in the semi-log plot is a robust
consequence of the momentum diffusion 
Eq.~(\ref{eq:smear}) by the multiple scattering,
irrespective of the detailed form of the transition probability.
Therefore the Pb--Pb data cannot be explained as the momentum diffusion effect
of the quark pair due to the multiple scattering
before forming a physical resonance.
 
We like to comment that 
in the  high-energy collisions
the transverse momentum diffusion in two dimensions
might be more appropriate than the three-dimensional one,
as in the Glauber approach.
Then the $L$ dependence of the suppression would change from
$L^{-3/2}$ to $L^{-1}$.

The absorption of a hadronic state by a power in $L$ is similar to
the result of the color transparency at high energies
predicted in many theoretical studies\cite{color,FM02}.
In the model studied here, however, the color degrees of freedom 
of the pair states are not explicitly treated in the transition probability
nor any interference effect.
Hence the relation to the
color transparency effect is unclear within this model. 
At the collider energies like RHIC and LHC the coherence effect becomes
more important and the quantum description of the
resonance production at parton level should be more elaborated.

The uniform shift of the relative momentum is the result of
the selective sum of higher twists at the leading order of $\alpha_s$
and the nuclear size\cite{F02}.
From the view of our model, 
it may be worthwhile to investigate the smearing effect due to
the correction terms in the perturbative QCD calculation.

In conclusion,  the suppression of the J/$\psi$ cross section
in $AB$ collisions except the Pb-Pb at SPS energy
can be described by the momentum diffusion
by the multiple scattering of the $c \bar c$ pair before the 
resonance formation. The suppression of the model is insufficient
to explain the J/$\psi$ yield observed 
in Pb-Pb collisions at SPS energy, which is therefore left anomalous
in this treatment.

The author is grateful to the referee for pointing out
the confusion in the original manuscript.
He also thanks T.~Matsui and M.~Ohtani for useful comments.
This work is supported in part by
 the Grants-in-Aid for Scientific Research of
Monka-sho (13440067),
and in part by Saneyoshi Scholarship Foundation.

\end{document}